\documentstyle[12pt]{article}
\input{psfig}
\newcommand{\pom}{I\!\! P}
\begin{document}
\vspace*{-6ex}
{\hfill
\fbox{Rockefeller University Report:  RU 97/E-29}}
\vglue 1.5cm
\begin{center}

{\large\bf
Factorization and Scaling in Hard Diffraction\footnote{\normalsize 
Presented at the Fifth International Workshop on Deep 
Inelastic Scattering and QCD, DIS97, Chicago, U.S.A, 14-18 April 1997.}}

\vglue 0.3cm
{K.  GOULIANOS\\ }
(dino@physics.rockefeller.edu)
\vglue 0.2 cm
\baselineskip=13pt
{\em The Rockefeller University, 1230 York Avenue\\
 New York, NY 10021, USA\\}
\vskip0.25in

\today

\end{center}

\centerline{ABSTRACT}

{\rightskip=10pc
\leftskip=10pc
We compare results on diffractive $W$-boson production
at the Tevatron with predictions based on the diffractive
structure function measured in deep inelastic scattering
at HERA assuming (a) conventional factorization or (b) hard
factorization combined with a rapidity gap
distribution scaled to the total gap probability.
We find that conventional factorization fails, while the
scaling prediction agrees with the data.
}
\section{Introduction}
Hard diffraction is defined as the class of hadronic diffractive processes that
incorporate a hard scattering.  
Recently, CDF reported results on diffractive $W$-boson~\cite{CDFW}
and dijet~\cite{CDFJ} production in 
$\bar p p$ collisions at $\sqrt{s}=1.8$ TeV 
at the Tevatron. In this paper, 
we compare the CDF results with predictions based on the diffractive 
structure function (SF) of the proton measured in $e^+p\rightarrow 
e^++[\gamma^*p\rightarrow Xp]$ deep inelastic 
scattering (DIS) at HERA~\cite{ZEUS,H1}.
Two such comparisons are made, one assuming conventional factorization
and the other assuming that the rapidity gap (RG) dependence of the 
diffractive SF scales to the total RG probability. 

Our predictions for the Tevatron can be obtained directly from the HERA 
diffractive structure function without reference to a ``pomeron flux" 
or even to the pomeron. However, to relate them to predictions based on
the standard and renormalized~\cite{R} pomeron flux factors, 
it is  useful to first introduce the pomeron flux language~\cite{IS}. 

In both the CDF and HERA cases,
the hard scattering involves
a parton from the pomeron, $\pom$, which is presumed to be ``emitted" by the
proton.  In $\bar pp\rightarrow Xp$, the $\pom$-parton interacts with a 
$\bar p$-parton producing a $W$ or a dijet, while in DIS the $\gamma^*$ is
absorbed by a quark in the $\pom$.
The proton emitting the pomeron remains intact, carrying a fraction $x$ of its
initial momentum; the remaining fraction, $\xi=1-x$, is carried by the pomeron.
Because of the colorless nature of the $\pom$
($\pom$ has the quantum numbers of the vacuum), a rapidity gap 
(absence of particles) occurs  between the final (surviving) 
proton and the particles in the system $X$.  The gap, whose nominal width is
$\Delta y=\ln\frac{1}{\xi}$, provides a characteristic signature for diffraction and is
used to identify  (``tag") diffractive events.  
The CDF and HERA data discussed here were selected using a rapidity
gap tag and are integrated over the transverse momentum squared, $t$, of the 
leading proton.

The differential hard diffraction cross section 
$d^4\sigma_{hard}/d\xi dtdQ^2d\beta$, where $\beta$ is the momentum 
fraction of the parton in the pomeron that participates in the hard scattering,
is the product of the hard $\pom-\bar p$  (or
$\gamma^*-\pom$ in DIS) cross section,  
which depends on the pomeron structure function,
and a ``pomeron flux factor", $f_{\pom/p}(\xi ,t)$, which is the probability
density of pomerons ``carried"  by the proton.  
In Regge theory, the flux factor has the form
\begin{equation}
f_{\pom /p}(\xi,t)\equiv \frac{{\beta_{\pom p}^2(t)}}{16\pi}\;\xi^{1-2\alpha(t)}
=\frac{{\beta_{\pom p}^2(0)}}{16\pi}\;\xi^{1-2\alpha(t)}F^2(t)
=K\;\xi^{1-2\alpha(t)}F^2(t)
\label{flux}
\end{equation}
\noindent where $\alpha(t)=1+\epsilon+\alpha' t$ is 
the pomeron trajectory, 
$\beta_{\pom p}(t)$ is the  coupling of the pomeron to the proton,
and $F(t)$ the proton form factor. Following Ref.~\cite{R}, we will use 
$\epsilon=0.115$, $\alpha'=0.26$ GeV$^{-2}$, 
$K=0.73$ GeV$^{-2}$ and $F^2(t)\approx e^{4.6t}$
(valid at small-$|t|$).

Assuming Regge factorization, the diffractive SF 
is expected to be the product of the pomeron flux times the pomeron SF:
\begin{equation}
F^{D(4)}_2(\xi,t,Q^2,\beta)
\equiv f_{\pom /p}(\xi,t)\cdot F_2^{\pom}(Q^2,\beta)
=\frac{Ke^{b(\xi)t}}{\xi^{1+2\epsilon}}
\cdot F_2^{\pom}(Q^2,\beta)
\label{F2D4}
\end{equation}

\noindent where $b(\xi)=4.6+2\alpha'\ln\frac{1}{\xi}$.  
Below, after reviewing the CDF results and standard pomeron flux predictions,
we present the diffractive structure function measured at HERA, use it to
predict the CDF results assuming factorization or RG scaling, 
relate the RG scaling to the renormalized pomeron flux of Ref.~\cite{R}, 
and draw conclusions on factorization and scaling in diffraction.
\section{Diffractive $W$ and dijet production}
The CDF Collaboration reported diffractive 
to non-diffractive ratios for 
$W$~\cite{CDFW} and dijet~\cite{CDFJ}  production, as well as 
Monte Carlo predictions based on POMPYT for diffractive and PYTHIA for 
non-diffractive events. The diffractive events were generated using
the standard pomeron flux and 
a hard pomeron structure of the form
$f^{\pom}(\beta)=(f_q+f_g)\cdot [6\beta(1-\beta)]$.
The predicted rates depend on the product of the 
quark (gluon) fraction of the pomeron, $f_q\; (f_g)$, 
and on the normalization of the 
pomeron flux factor. 
The gluon fraction of the pomeron can be determined 
from the ratio of the $W$ to the dijet measured rates independent of the 
flux normalization or of the validity of the momentum sum rule for the pomeron.
This is possible due to the different sensitivity of the $W$ and dijet 
production rates to the quark and gluon content of the pomeron.
Thus, any deviation from unity found in the  ratio 
of measured to predicted rates, $D$, can be attributed either to 
a discrepancy in the flux 
normalization or to a 
failure of the pomeron momentum sum rule (defined as $f_q+f_g=1$).
Comparison of the measured $W$ rate with the rate predicted 
from the diffractive structure function 
measured in DIS provides a direct test of conventional factorization.  

The diffractive to non-diffractive ratios measured by CDF are:

\begin{center}
$R_W=[1.15\pm 0.51(stat)\pm 0.20(syst)]\%=(0.115\pm 0.55)\%$\\
$(\xi<0.1)$
\end{center}
\begin{center}
$R_{JJ}=[0.75\pm 0.05(stat)\pm 0.09 (syst)]\%=(0.75\pm 0.10)\%$\\
($E_T^{jet}>20$ GeV, $|\eta|^{jet}>1.8$, $\eta_1\eta_2>0$, $\xi<0.1$)
\end{center}
The POMPYT standard flux predictions are $R _W^{MC}=16\%$ (1.1\%) for a three 
quark-flavor (full-gluon) pomeron, and $R_{JJ}^{MC}=2\%$ (5\%) 
for a full-quark (full-gluon) pomeron structure. From these predictions 
and the measured rates, CDF derived the gluon fraction, $f_g$, 
and the pomeron flux/momentum discrepancy factor, $D$:
$$f_g=0.7\pm 0.02\;\;\;\;\;D=0.18\pm 0.04$$
\section{The diffractive structure function}
Both the ZEUS~\cite{ZEUS} and H1~\cite{H1} Collaborations find that 
in the region $8.5<Q^2<65$ GeV$^2$ the  
integral of the $F_2^{D(4)}$ structure function over $t$ 
has a form similar to that of Eq.~\ref{F2D4}, namely 
\begin{equation}
F_2^{D(3)}(\xi,Q^2,\beta)=
\displaystyle{\int_{-{\infty}}^{t_{min}}}F_2^{D(4)}(\xi,t,Q^2,\beta)\,dt=
\frac{1}{\xi^{1+n}}\cdot A(Q^2,\beta)
\label{F2D3}
\end{equation}
where $n\sim 0.2-0.3$. 
To simplify numerical comparisons with Eq.~\ref{F2D4} we will use 
$n=2\epsilon=0.23$.  The term $A(Q^2,\beta)$ is rather flat in $\beta$ 
and increases slowly with $Q^2$. 
Its average value is represented well by the value at $Q^2\sim 20$ GeV$^2$, 
$A(Q^2,\beta)|_{Q^2=20}
\approx 0.009$. To facilitate comparison with the CDF predictions based 
on the pomeron flux we will use $A(Q^2,\beta)|_{Q^2=20}=
0.009[6\beta(1-\beta)]$. 

\subsection{Factorization}
To calculate $R_W$ from $F_2^{D(3)}$, we first establish the correspondence 
with the pomeron flux, so that we may make use of the MC 
results of CDF. Following CDF, we set 
$$F_2^{\pom}(Q^2,\beta)=\frac{2}{9} f_q \cdot 6\beta(1-\beta)$$
\noindent where the factor 2/9 is the average quark charge for three quark 
flavors (3f) in the pomeron and $f_q$ is the quark fraction of the pomeron. 
Using this form for $F_2^{\pom}$ 
and equating the integral over $t$ of Eq.~\ref{F2D4}   
with the structure function (SF) measured at HERA, we obtain
\begin{equation}
\frac{K}{\xi^{1+2\epsilon}}\cdot  \frac{1}{7.5} \cdot 
\left[\frac{2}{9}f_q\cdot
6\beta (1-\beta)\right]=\frac{0.009}{\xi^{1.23}} 
[6\beta(1-\beta)]
\label{F2D320}
\end{equation}
where we have used the average $t$-slope of 7.5 GeV$^{-2}$ in carrying 
out the integration over $t$. With the standard flux normalization, 
$K=0.73$ GeV$^{-2}$, the quark fraction turns out 
to be $f_q=0.42$, which multiplied by the MC prediction of 16\% for a 
full 3f-quark pomeron yields $R_W^{SF}=6.7\%$. 
This value is 5.8 times larger than the measured value of $R_W=1.15\%$. 
Thus, conventional factorization breaks down~\cite{Whitmore}.
\subsection{Scaling}
Let us now assume that the RG distribution scales to the total gap 
probability and rewrite the $F_2^{D(3)}$ to reflect such scaling.
We note that for fixed $Q^2$ and $\beta$ the kinematically allowed $\xi$-limits
are $\xi_{min}=Q^2/\beta s$ and $\xi_{max}\approx 0.1$ (the coherence 
limit)~\cite{R}. Integrating the factor $\xi^{-1.23}$ between these limits 
yields
\begin{equation}
N(Q^2,\beta,s)\equiv N(\xi_{min})= 
\displaystyle{\int_{\xi_{min}}^{0.1}}
\frac{1}{\xi^{1.23}}d\xi=\frac{1}{0.23}
\left[\left(\frac{\beta s}{Q^2}\right)^{0.23}-1.7\right]
\label{N}
\end{equation}
The scaled SF can now be written as
\begin{equation}
F_2^{D(3)}(\xi,Q^2,\beta)|_{scaled}\approx
\frac{0.009}{\xi^{1.23}}\cdot
[6\beta(1-\beta)]\cdot \frac{C}
{N(Q^2,\beta,s)}
\label{F2D3R}
\end{equation}
where the constant $C$ is chosen to be $C=N(20,0.5,300^2)=18.3$
so that the scaled SF for $Q^2=20$ and $\beta=0.5$ is the same as 
the unscaled SF.
Through $N(Q^2,\beta,s)$, the scaled SF 
acquires an additional $\beta$ dependence and a 
$Q^2$ dependence. The extra $\beta$ 
dependence helps flatten the $\beta(1-\beta)$ distribution at small $\beta$ 
and make it more similar to that measured at HERA. The acquired $Q^2$ 
dependence is very close to that observed at HERA. 
Thus, in the pomeron flux language, it is the flux 
at fixed $\xi$ that increases with $Q^2$ while the pomeron 
structure remains relatively unchanged. 
For a more detailed discussion see Ref.~\cite{R}.  

To use the SF of Eq.~\ref{F2D3R} at the Tevatron, 
one simply has to evaluate $N(Q^2,\beta,s)$ for $\sqrt{s}=1800$ GeV and 
$Q^2=M_0^2=1.5$ GeV$^2$, where $M_0$ is the effective diffractive 
threshold for $\bar pp\rightarrow Xp$.  
For $\beta=0.5$, $N(M_0^2,\beta,s)_{TEV}=98.8$.
This value is larger than the corresponding HERA value of 18.3
by a factor of 5.4. Thus, the $R_W^{SF}=6.7\%$ has to be multiplied by 
the scaling factor $D_{scale}=1/5.4=0.19$, yielding 1.24\%. 
The value of 0.19 agrees with 
the standard flux discrepancy factor 
of $0.18\pm 0.04$ reported by CDF.
The scaled SF prediction for the diffractive to non-diffractive $W$ 
production ratio, 1.24\%, is in
excellent agreement with the measured value of $(1.15\pm 0.55)$\%. 
Note that using the measured SF at a $Q^2$ value other than $Q^2=20$  
would yield a slightly different prediction for $R_W$, but the 
prediction obtained with the scaled flux would remain the same. 
\newpage
\vglue 0.2cm
\centerline{\psfig{figure=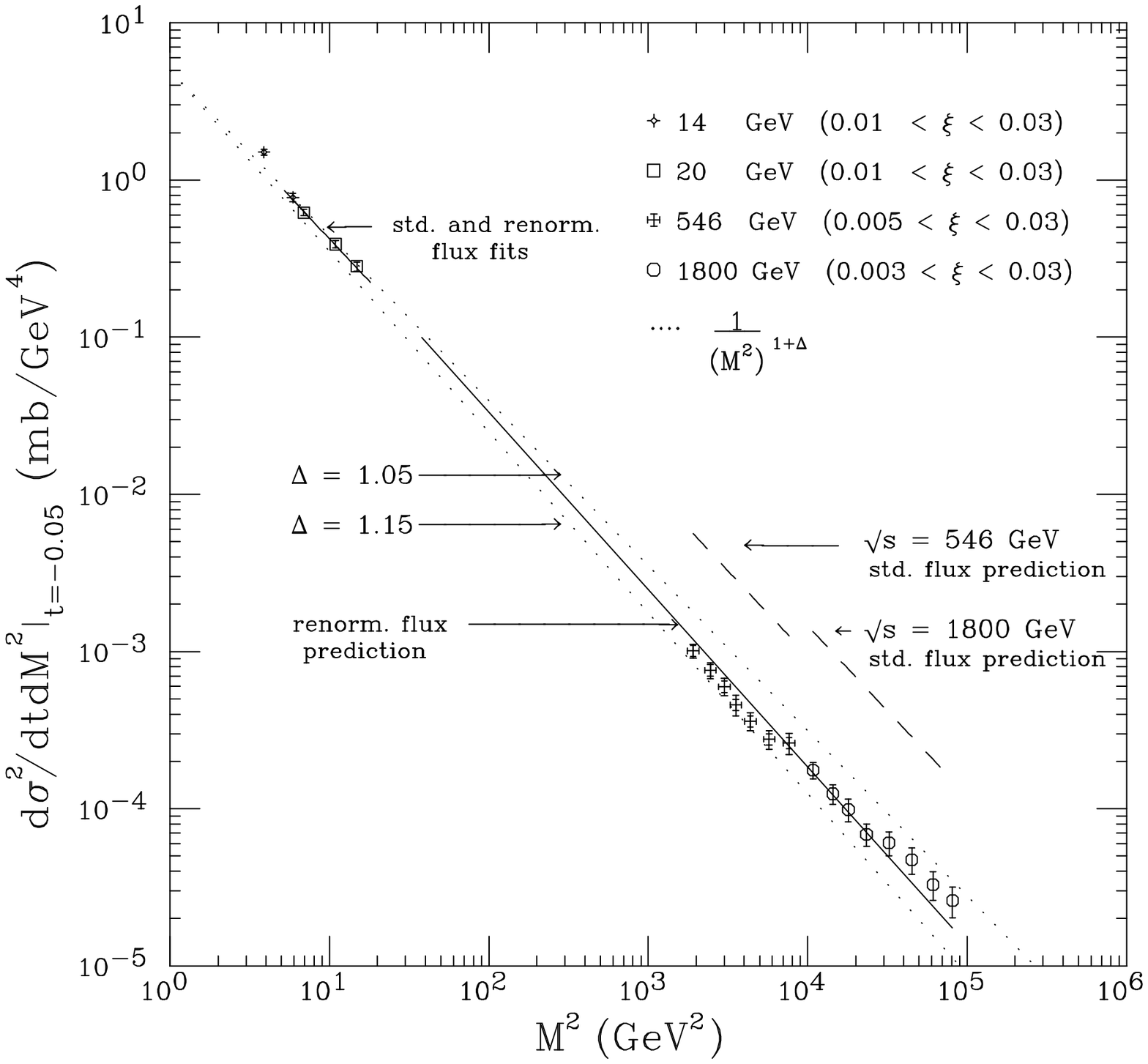,width=4.5 in}}
\vglue -0.25in
Figure 1: Diffractive differential cross sections 
$d^2\sigma/dtdM^2|_{t=-0.05}$ as a 
function of $M^2$ for $pp$ 
data~\cite{C} at $\sqrt{s}=14$ and 20 GeV and 
$\bar pp$ data~\cite{CDFD} at $\sqrt{s}$=546 and 1800 GeV shown along 
with the standard and renormalized flux predictions. Both sets of 
predictions are
normalized to the 20 GeV data points. 
\section{Scaling and the renormalized pomeron flux}
The scaling of the diffractive SF to the total gap probability is 
equivalent to the renormalized flux hypothesis of Ref.~\cite{R}.  
When correctly applied to hard processes, i.e. using $\xi_{min}=M_0^2/\beta s$
in evaluating the normalization factor $N(\xi_{min})$  (where $M_0$ is the 
effective diffractive mass threshold), the renormalized flux gives 
the same results as the scaled structure function. 
Thus, the factorization breakdown 
observed between DIS and hard diffractive production at the Tevatron is 
traced back to the breakdown of factorization in soft diffraction. 

The pomeron flux scaling implies that 
the soft diffractive cross section $d^2\sigma/dM^2dt|_{t=0}$ is approximately 
independent of $s$, contrary to the expectation of an $\sim s^{2\epsilon}$ 
dependence from the triple-pomeron amplitude.  Figure~1, which 
shows $d^2\sigma/dM^2dt|_{t=-0.05}$ for $pp/\bar pp$ data at 
different $s$-values~\cite{C,CDFD}, 
confirms the scaling of the $M^2$-distribution with $s$~\cite{M}.
\section{Conclusions}
We have compared the CDF results on diffractive $W$ production 
with predictions using the structure function  
measured at HERA. We find that 
conventional factorization fails, but a SF in which the gap probability 
distribution is scaled to the total gap probability yields predictions which 
are in excellent agreement with the data. 
Through the connection between  the scaled SF
and the renormalized pomeron flux, we conclude that the breakdown of 
factorization of the diffractive SF of the proton 
is due to the breakdown of the Regge factorization already observed in soft 
diffraction and, therefore, that factorization {\em 
of the pomeron structure function} in hard processes 
still holds. Finally, we have shown that the approximate
scaling of $d^2\sigma/dtdM^2|_{t=0}$ with $s$ implied by flux 
renormalization is confirmed by the data.

\end{document}